\documentclass{aa}

\usepackage{epsfig}

\begin{document}

\thesaurus{}
\titlerunning{Signature of the galactic bar}
\title{Evidence for a signature of the galactic bar in the solar neighbourhood
\thanks{Based on data from the ESA HIPPARCOS satellite, on photometric data
collected at the Swiss telescope at La Silla and on data collected with
CORAVELs at the Haute Provence Observatory (France) and at ESO, La Silla Chile}}
\author{D. Raboud \and M. Grenon \and L. Martinet \and R. Fux \and S. Udry}

\institute{Observatoire de Genève, CH-1290 Sauverny}

\date{Received date; accepted date}

\maketitle

\begin{abstract}

Using available kinematical data for a subsample of NLTT stars from the HIPPARCOS mission
we confirm the existence of a previously reported local anomaly in the $(u,v)$ plane:
the mean motion $\overline{u}$ for old disc stars, with $v < -30$ km s$^{-1}$, is largely
positive ($+19 \pm 9$ km s$^{-1}$ w.r.t. the Galactic Center).

With the use of the newest global self-consistent numerical models of our Galaxy
(Fux \cite{fux}), we show that a bar could be responsible for this observed velocity anomaly.
A fraction of our stars have bar perturbed ``hot'' orbits, allowing them to
erratically wander from inside the bar to regions outside the corotation, in particular
through the solar neighbourhood.


\end{abstract}

\section{Introduction}

The kinematical properties of disc stars in the solar neighbourhood have been
extensively studied in the past. Velocity dispersion increase with age and
vertex deviation have been well known facts for a long time and have been
recently reconfirmed  (Dehnen \& Binney \cite{dehnen}).

A feature of local stellar motions which practically passed unnoticed is
a global asymmetry of the star distribution in the $(u,v)$ plane of velocities
relative to the Sun \footnote{$u$ positive in the anti-center direction, $v$ in the
rotation direction}, in addition to the vertex deviation which essentially concerns
stars of small epicyclic energies: the mean \textit{u}-velocity ($\overline{u}$) of
the old disc stars, as inferred from various samples, appears significantly
different from zero even if we correct it for the solar motion, contrarily to the
expectation for a stationary axisymmetric galaxy.
This anomaly was already apparent with the Gliese \& Jahreiss (\cite{gliese})
and the Woolley (\cite{woolley}) space-velocities of nearby stars and could be inferred from
the positions of Eggen's old disc stellar groups in the $(u,v)$ plane 
(Eggen \cite{eggen} and references therein). Mayor (\cite{mayor1}) emphasized a 
significant excess $\overline{u}(h) > 0$ ($h$ = angular momentum) for stars having
a mean asymmetrical drift $\langle S \rangle\ \approx 20$ - 30 km s$^{-1}$. 
Nevertheless, a quantitative interpretation of the origin of this phenomenon has 
never been given until now.

Two new facts allow us to go more deeply into this question. At first, after the
de Vaucouleurs (\cite{devaucoul}) presumption, series of recent more or less direct 
observational evidences indicate that our Galaxy is definitely barred 
(see e.g. Kuijken \cite{kuijken} for a review). Fux (\cite{fux}) realised global
self-consistent numerical simulations of our Galaxy which all develop a long-lasting bar.

Secondly, Grenon observed a large sample of 5443 NLTT (New Luyten's Two Tenths, i.e.
$\mu > 0.18 \pm 0.02$ arcsec yr$^{-1}$) stars for which space velocities
and partly chemical compositions are available. The sample will be described in details
in Sect. 2. Its size allows us to deal with outstanding questions of galactic structure
and evolution.

In the present letter, taking benefit of these progresses, we aim to examine
whether the bar could be responsible for the kinematical anomaly mentioned above.

\section{Observational data}

The nearby star samples are notoriously poor in halo and very old disc stars. In
order to test the effect of a bar in the inner Galaxy, we must consider stars moving 
on eccentric orbits bringing them in the bar proximity. The sample used here is an 
extension of the Gliese's and Woolley's catalogues towards larger volume and space
velocities at the expense of the completeness in low space-velocity stars. It is a
subset of a vast programme prepared for the HIPPARCOS mission. The initial set of 
10047 stars was formed of all stars from Luyten's NLTT Catalogue with $m_{R} < 11.5$ if of color class
from a to k-m, or $m_{R} < 12.5$ if  the class was m or m+. According to internal 
priorities favouring the selection of parallaxe stars, namely those within 100 pc, and
to satellite observing possibilities, 7824 stars have been included in the HIPPARCOS
programme.

Since the sample is absolute magnitude limited with $M_{\mbox{v}} < 6.5$ at 100
pc, i.e. well below the turn-off luminosity, it shows no bias against metal-rich
and super metal-rich (SMR) stars (Grenon \cite{grenon2}). It is also proper-motion 
limited and hence tangential velocity limited, i.e. $V_{t} > 0.85\cdot d$ km s$^{-1}$
with distance $d$ in [pc]. 

Our sky coverage $\delta < +10\degr$ includes in particular a complete south galactic cap
defined by $b < -53\degr$.                                           
For the HIPPARCOS mission preparation and for the obtention of physical parameters
from photometry, all programme stars south of $\delta=+10\degr$ were observed at the
Swiss telescope at La Silla, from 1981 on. A total of 39435 measurements were
accumulated for 5443 different stars.
Calibrations of the Geneva photometric system and susbsequent revisions by
Grenon (\cite{grenon1})
were used to derive $T_{\small{\mbox{eff}}}$, $M_{v}$ and $[M/H]$, with precisions 
varying with the temperature and gravity ranges. For late G dwarfs, the internal errors
are of 20-40 K on $T_{\small{\mbox{eff}}}$, 0.03-0.05 dex on $[M/H]$ and 0.15 on $M_{v}$,
for single stars. Stars later than K3V have no $[M/H]$ estimate.

\begin{figure}[t]
\centerline{\psfig{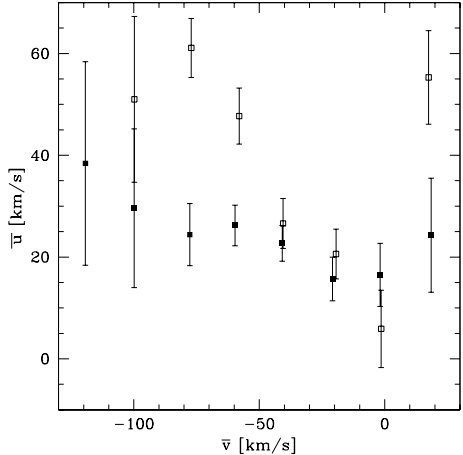}}
\caption{Observed $\overline{u}$ as
a function of the $\overline{v}$ (w.r.t. the Sun) velocities for disc stars
with $\pi < 40$ mas, at South Galactic Pole (filled squares, 673 stars) and 
towards the anti-rotation (open squares, 462 stars with $l$ between
240\degr\, and 300\degr\, and $|b| < 40\degr$).}
\label{fig:Ubh}
\end{figure}

Radial velocities were obtained for all programme stars by Grenon with CORAVEL at 
Haute-Provence Observatory in the declination zone $\delta=$ -20$\degr$ to +10$\degr$
and as part of a dedicated ESO Key Programme (Mayor et al. \cite{mayor2}) south
of  $\delta=$ -20$\degr$. A few velocities were taken from literature, in particular 
from the Carney et al. (\cite{carney}) and from the Barbier-Brossat et al. (\cite{barbier}) 
catalogues. Binaries were identified either from the radial velocity changes or the
comparison between trigonometric and photometric parallaxes, or detected by HIPPARCOS 

For the present study, stars were kept if their HIPPARCOS and photometric parallaxes 
were larger than 10 mas. Moreover, uncertain parallaxes with standard errors $ >
5$~mas
were rejected. In our final sample, space velocities are available for 4143 stars and
overall metallicities for 2619 of them.

\section{Kinematical properties of the sample}

The old disc stars of our sample show largely positive mean $\overline{u}$
motions, confirming the trend already apparent in the earlier results recalled
in the Introduction.

At given $v < -30$ km s$^{-1}$, there is no bias in favour of inward or outward
motions when stars are observed in the south galactic cap or towards the
anti-rotation where $V_{t}=\sqrt{u^{2}+w^{2}}$. With increasing distances the
fraction of small $|u|$ decreases in favour of large $u+$, $u-$ velocities. When
old disc stars are selected in the $\pi$ range from 10 to 40~mas, the unbalance
between $u+$ and $u-$ becomes stricking, see Fig.~\ref{fig:Ubh}, and reaches up
to $+50$ km s$^{-1}$ towards the anti-rotation. When subsamples are complete
according to $v$ components (e.g. all stars with $v < -50$ km s$^{-1}$ are
retained), the $u$ anomaly is correctly estimated with values within the range
$+29 \pm 2$ km s$^{-1}$ w.r.t. the Sun, i.e.
$+19 \pm 9$ km s$^{-1}$ w.r.t. the Galactic Center (GC) after correction for
local motions inferred from Kuijken \& Tremaine (\cite{KuijTre}).

\begin{figure}[t]
\centerline{\psfig{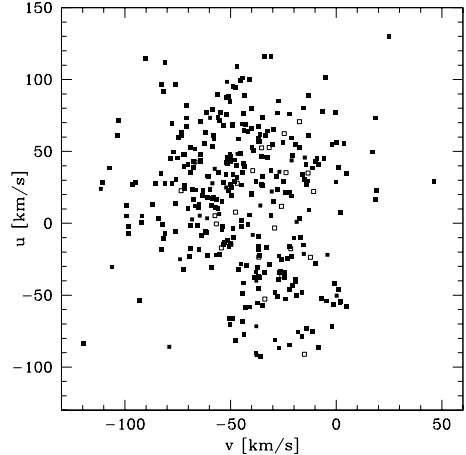}}
\caption{Observed $(u,v)$ plane (w.r.t. the Sun) for metal-rich stars with
$[M/H]$ between 0.25 and 0.65, without Hyades group members. Filled squares stand
for single stars and open squares for multiple ones.}
\label{fig:planUV}
\end{figure}

Stars with metallicities atypical for the young local disc, i.e. with $[M/H] >
0.25$ or $ < -0.30$, form the best test sample to investigate large scale
perturbations of stellar orbits. The metal poor group should contain stars born
inside and outside the solar orbit, whereas the metal-rich group should consist
of stars born uniquely within the solar orbit. It is therefore well suited to
test the orbital anomalies in the inner disc. Its $(u,v)$ plane is displayed in
Fig. \ref{fig:planUV}. 

\begin{figure}[t]
\centerline{\psfig{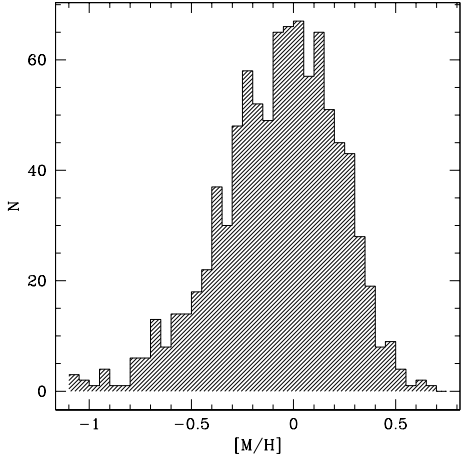}}
\caption[]{Metallicity distribution for NLTT stars with $u~>~10$~km~s$^{-1}$
and $-130 < v < -30$ km s$^{-1}$. This distribution is similar to the
one observed for stars of the galactic bulge (see Fig. 17 of McWilliam
\& Rich (\cite{McWilliam})).}
\label{fig:mhhsg}
\end{figure}

\begin{figure}[t]
\centerline{\psfig{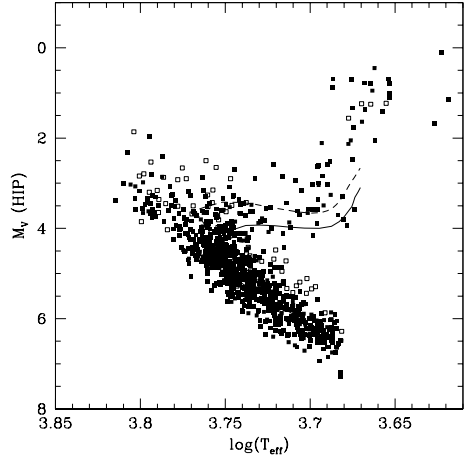}}
\caption[]{HR diagram for stars with $[M/H] > -0.55$. The solid
and dashed lines
correspond to isochrones of log t = 10.0 and 9.8 respectively. The filled
squares stand for single stars and the open squares for multiple ones.}
\label{fig:HR}
\end{figure}

If we consider the $(u,v)$ plane for all stars
with known $[M/H]$, the area $u > 10$ km s$^{-1}$ and $ -130 < v < -30$ km~s$^{-1}$
contains not only metal-rich stars but a sample of old disc stars with metallicities
covering the whole range of $[M/H]$ observed in the old disc population from $\sim -0.6$
dex to $+0.6$ dex and showing a peak at solar metallicity (Fig. \ref{fig:mhhsg}). The $[M/H]$
distribution is essentially the same as that observed in situ in the bulge by McWilliam
\& Rich (\cite{McWilliam}), an indication of
a common origin for both populations. The ages are also heterogeneous with a range
between 6 to 10 Gyr (Fig. \ref{fig:HR}). 

The $\overline{u}$ anomaly has hence to be interpreted as a common response
of old stars to secular gravitational sollicitations rather than as a memory
of initial conditions.

\section{Stellar response to a bar}

In this section we compare the observational features mentioned above with the
local kinematics of old disc particles inferred from simulation with global
self-consistent 3-D numerical models of our Galaxy developed by Fux
(\cite{fux}). These models have
a bar axis ratio $b/a=0.5\pm0.1$ and a bar pattern speed $\Omega_{p}=50\pm10$
km~s$^{-1}$ corresponding to a corotation radius of $4.3\pm0.5$ kpc. The Fux model m08
at $t$=3.2 Gyr presents the best agreement with the different local and global
observational constraints considered by Fux.

\begin{figure}[t]
\centerline{\psfig{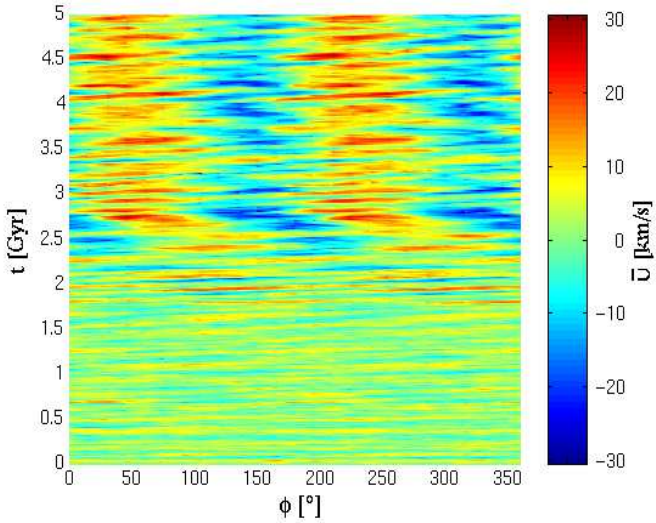}}
\caption[]{Time behaviour of $\overline{u}(\phi)$
over 5 Gyr, for stars observed within a torus between $\tilde{R}$ = 7.5 and
8.5 (in the initial units of the simulation m08, Fux \cite{fux}).}
\label{fig:ubphi}
\end{figure}

In this m08 simulation we follow the mean radial motion $\overline{u}$ of
particles w.r.t. the GC as a function of time, across a toroidal region of 1 kpc
diameter at a distance $R=R_{0}=8$ Kpc from the GC. In the considered models, built
with particules which essentially represent the old disc, the global
deviation with respect to axisymmetry is practically negligible before 2 Gyr 
(Fig.~\ref{fig:ubphi}). The bar, if it exists, could be at most
confined to a very small region around the center. Some transitory spiral
density waves are observed, but $\overline{u}$ is never very different from zero
during this early period. However, as soon as the bar is stabilized (for $t >
2.9$ Gyr),
$\overline{u}$ differs significantly from 0 and depends on $\phi$, the angle between
the bar and the line Sun-galactic center.

\begin{figure}[t]
\centerline{\psfig{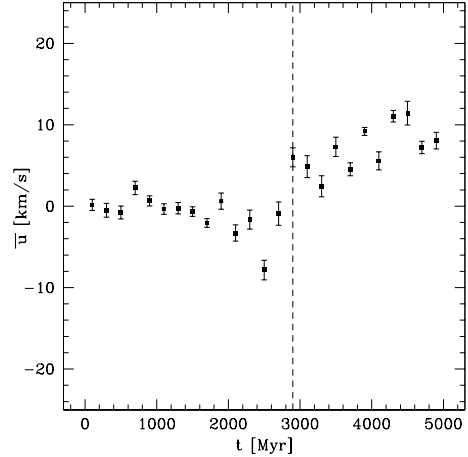}}
\caption[]{Time behaviour of $\overline{u}(\phi=30\degr)$
over 5 Gyr, averaged in bins of 200 Myr, for stars located within a torus
between $\tilde{R}$ = 7.5 and 8.5 (in the initial units of the simulation m08,
Fux \cite{fux}). In this simulation the bar is completely settled at the right
of the vertical dashed line. This graphic is merely a smoothed cut through Fig.
\ref{fig:ubphi}, for $\phi=30\degr$.}
\label{fig:ut}
\end{figure}

The bar produces a sine-like behaviour in $\phi$ for the mean velocity
$\overline{u}$ (note that the streaming motion $\overline{u}$ is zero when
integrated over $\phi$ for several bar rotational periods, even for stochastic
orbits). The mean amplitude of $\overline{u}(\phi)$ is
$\sim 17$ km~s$^{-1}$, but $|\overline{u}|$ values as high as 28 km~s$^{-1}$
are temporarily obtained.  For example, around $t = 4.53$ Gyr and
$\phi =$ 30\degr\, (which is the angle corresponding to the best statistical agreement
of the models with the COBE/DIRBE data),
$\overline{u}$ of the order of 20 km s$^{-1}$ are obtained, in agreement with the
observational results reported in Sect. 3. Figure~\ref{fig:ut} is a slice
cutting of Fig. \ref{fig:ubphi} for $\phi = 30$\degr. 
The values of $\overline{u}$ have been averaged within bins of 200 Myr width.
One obtains a significantly positive value of $\overline{u}$ when the bar is
stabilized.

The presence of a bar in our Galaxy implies characteristic orbital behaviours
of stars. Detailed discussions on the kind of trajectories
followed by stars in self-consistent barred potentials have been presented by
Sparke \& Sellwood (\cite{sparke}), Pfenniger \& Friedli (\cite{pfenniger}) and
Kaufmann \& Contopoulos (\cite{kaufmann}). Let us distinguish here two
categories of orbits: 1)~elongated orbits confined to the bar,
trapped about the long-axis $x_{1}$ family of periodic orbits (bar particles),
2)~``hot'' orbits
which essentially display a typical chaotic behaviour, erratically wandering between
regions inside the bar and outside corotation. Whereas orbits of kind 1)
have an Hamiltonian $H < H(L_{1,2})$, the ``hot'' orbits have $H > H(L_{1,2})$,
$H(L_{1,2})$ being the Hamiltonian at the Lagrange point $L_{1}$ and $L_{2}$ for
a zero velocity in the rotating frame.
Examples of these kinds of orbits can be seen for ex. in Fig. 15 of Sparke \& 
Sellwood (\cite{sparke}).

In the solar neighbourhood, we are not able to observe the first category of 
orbits as the Sun is outside corotation. But at the present time we observe near the
Sun some number of hot orbit stars which are able to visit our neighbourhood after
spending more or less time in the inner region of the bar.
In particular metal-rich old disc stars, which are typical contributors to the
kinematical anomaly according to Fig. \ref{fig:planUV}, follow such kind of orbits.
Those with $u > 0$ indeed have an Hamiltonian $H > H(L_{1,2})$.
As for the well known, but not considered in this letter, two peak (Hyades,
Sirius) distribution for stars with small epicyclic energy, the respective
influence of local spiral density waves (Mayor \cite{mayor1}) and of a bar
(Kalnajs \cite{kalnajs}) is not well established. It depends on the relative
strength of these perturbations.

\vspace{-0.2cm}
\section{Conclusion}

The present letter suggests an explanation for the kinematical anomaly observed in
the solar neighbourhood, characterized by a significant positive mean value of
the center-anticenter motions of stars with trailing azimuthal velocities with 
respect to the circular motion.

Our conclusions are the following:

1) Recent samples of old disc stars with $v < -30$ km~s$^{-1}$ confirm a mean positive
motion $\overline{u}$ of $19 \pm 9$ km s$^{-1}$ (w.r.t. the GC),
signature of an anomaly in the local kinematics with respect to a stationary
axisymmetric case.

2) Our numerical simulation shows that the bar in our Galaxy such as constrained by
various observational data (Fux \cite{fux}) produces a sine-like behaviour in $\phi$
for the mean radial velocity $\overline{u}$ w.r.t. the GC for old
disc particles. For the value of $\phi$ corresponding to
the presently Sun-galactic center angle, $\overline{u}$ is positive in agreement
with observations. These calculations
suggest that the local kinematic anomaly for the old disc could be a signature of the bar.

3) The bar is able to perturb orbits of stars either born or passing through
its region and wandering through the solar neighbourhood. For example, we calculate
that in a time interval of 5 Gyr a star with $(u,v,w)=(50,-50,0)$ km s$^{-1}$ at the
Sun can visit regions between $\sim3$ kpc and $\sim10$ kpc from the galactic center.
This range would be limited to 3 kpc in an axisymmetric galaxy.
``Hot'' orbits ($H > H(L_{1,2})$) detected in the solar neighbourhood,
populated by the metal-rich old stars mentioned above, are among these perturbed
orbits.

\end{document}